\documentclass[twocolumn,english,superscriptaddress,footinbib,bibnotes,reprint]{revtex4-1}

\usepackage[version=3]{mhchem}

\usepackage{graphicx}
\usepackage[usenames,dvipsnames]{color}
\usepackage{dcolumn}
\usepackage{bbm}
\usepackage{bm}
\usepackage{amsmath}
\usepackage{amssymb}
\usepackage{times}
\usepackage{placeins}
\usepackage{hyperref}
\hypersetup{colorlinks=false}

\newcommand{\NiRhO}{NiRh$_{\text 2}$O$_{\text 4}$}
\newcommand{\CoAlO}{CoAl$_{\text 2}$O$_{\text 4}$}
\newcommand{\FeScS}{FeSc$_{\text 2}$S$_{\text 4}$}
\newcommand{\MnScS}{MnSc$_{\text 2}$S$_{\text 4}$}

\begin{document}

\title{Quantum spin liquids in frustrated spin-1 diamond antiferromagnets}

\author{Finn Lasse Buessen}
\affiliation{Institute for Theoretical Physics, University of Cologne, 50937 Cologne, Germany}
\author{Max Hering}
\affiliation{Dahlem Center for Complex Quantum Systems and Institut f\"ur Theoretische Physik,
Freie Universit\"at Berlin, Arnimallee 14, 14195 Berlin, Germany}
\author{Johannes Reuther}
\affiliation{Dahlem Center for Complex Quantum Systems and Institut f\"ur Theoretische Physik,
Freie Universit\"at Berlin, Arnimallee 14, 14195 Berlin, Germany}
\affiliation{Helmholtz-Zentrum Berlin f\"ur Materialien und Energie, Hahn-Meitner-Platz 1, 14019 Berlin, Germany}
\author{Simon Trebst}
\affiliation{Institute for Theoretical Physics, University of Cologne, 50937 Cologne, Germany}

\date{\today}

\begin{abstract}
Motivated by the recent synthesis of the spin-1 A-site spinel \NiRhO, we investigate the classical to quantum crossover 
of a frustrated $J_1$-$J_2$ Heisenberg model on the diamond lattice upon varying the spin length $S$. 
Applying a recently developed pseudospin functional renormalization group (pf-FRG) approach for arbitrary spin-$S$ magnets, 
we find that systems with $S \geq 3/2$ reside in the classical regime where the low-temperature physics is dominated by the formation 
of coplanar spirals and a thermal (order-by-disorder) transition. 
For smaller local moments $S$=1 or $S$=1/2 we find that the system evades a thermal ordering transition and forms 
a quantum spiral spin liquid where the fluctuations are restricted to characteristic momentum-space surfaces. 
For the tetragonal phase of \NiRhO, a modified $J_1$-$J_2^-$-$J_2^\perp$ exchange model is found to 
favor a conventionally ordered N\'eel state (for arbitrary spin $S$) even in the presence of a strong local single-ion spin anisotropy
and it requires additional sources of frustration to explain the experimentally observed absence of a thermal ordering transition.
\end{abstract}

\maketitle

\begin{figure}[b]
  \centering
  \includegraphics[width=0.9\linewidth]{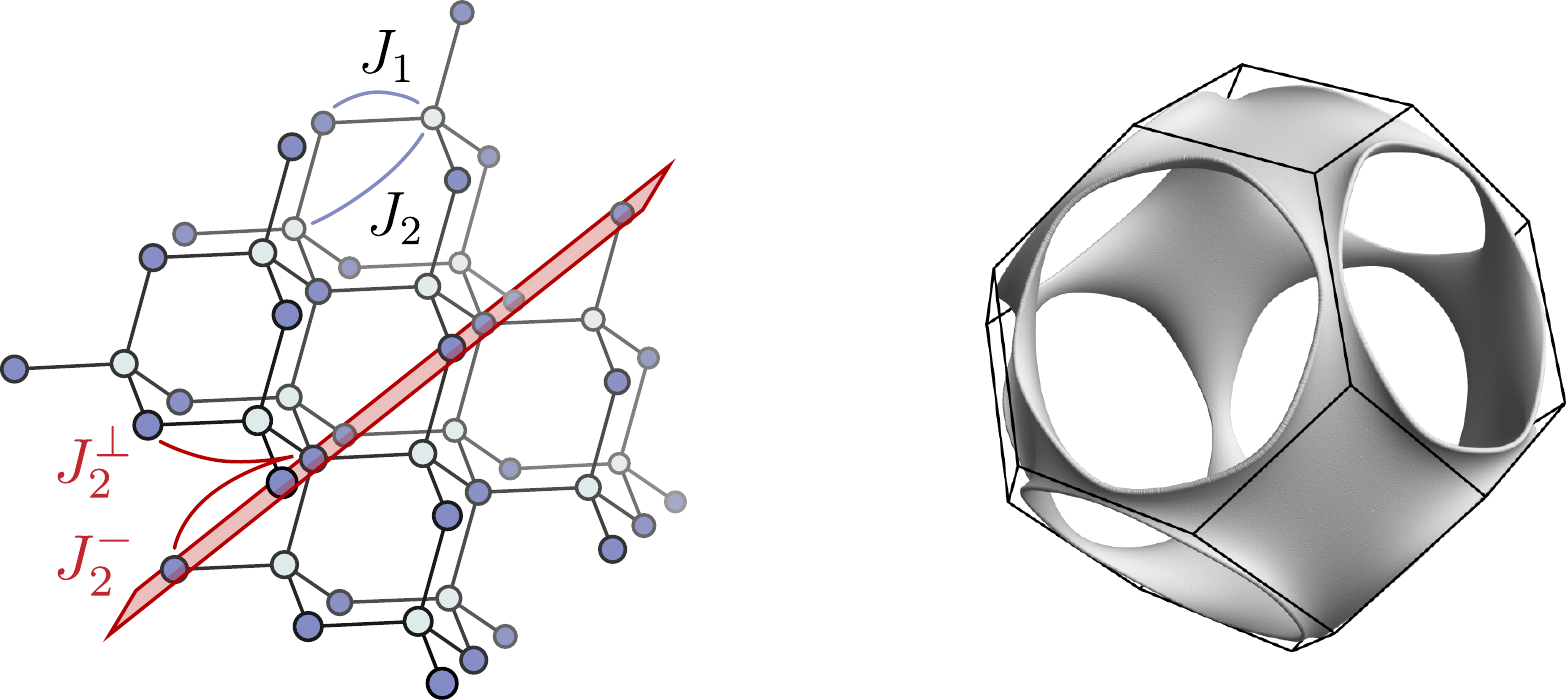}
  \caption{{\bf Frustrated diamond lattice antiferromagnet}. 
  		Left: Diamond lattice with nearest ($J_1$) and next-nearest neighbor coupling ($J_2$). 
  		A tetragonal distortion of the lattice along one spatial axis (orthogonal to the plane indicated in red) splits the 
		12 next-nearest neighbor couplings into a group of 4 in-plane terms ($J_2^-$) and 8 out-of-plane terms ($J_2^\perp$).  
		Right: Spin spiral surface for $J_2/|J_1| = 0.73$ plotted in the first Brillouin zone (solid lines).}
  \label{fig:lattice}
\end{figure}

In the field of frustrated magnetism, spinel compounds of the form AB$_2$X$_4$ (with X=O, Se, S) have long been appreciated as a 
source of novel physical phenomena \cite{Takagi2011}. 
B-site spinels with magnetic B ions and non-magnetic A ions, such as ACr$_2$O$_4$ or AV$_2$O$_4$ (with A=Mg, Zn, Cd),
realize pyrochlore antiferromagnets where geometric frustration mani\-fests itself in a vastly suppressed ordering 
temperature relative to the Curie-Weiss temperature. Conceptually, the pyrochlore Heisenberg antiferromagnet is a
paradigmatic example of a three-dimensional spin liquid  \cite{Balents2010,Savary2016}, 
in both its classical \cite{Moessner1998,Moessner1998b} and quantum \cite{Canals1998,Canals2000} variants.
A-site spinels, with non-magnetic B ions and magnetic A ions forming a diamond lattice, 
have caught broader attention some ten years ago with the synthesis of 
\MnScS~\cite{Fritsch2004}, 
\FeScS~\cite{Fritsch2004}, 
and \CoAlO~\cite{Tristan2005,Suzuki2007}
that, similar to the B-site spinels, exhibit a dramatic suppression of their ordering temperature. 
At first sight counterintuitive due to the unfrustrated nature of the diamond lattice, it was conceptualized \cite{Bergman2007}
that a sizable next-nearest neighbor coupling (connecting spins on the fcc sublattices of the diamond lattice) induces
strong geometric frustration. Indeed it could be shown that the classical Heisenberg model with both nearest and 
next-nearest neighbor exchange
\begin{equation}
	\mathcal{H} = J_1 \sum_{\langle i,j \rangle} {\bf S}_i {\bf S}_j + J_2 \sum_{\langle \langle i,j \rangle \rangle} {\bf S}_i {\bf S}_j 
	\,,
	\label{eq:heisenberg-J1J2}
\end{equation}
exhibits highly-degenerate coplanar spin spiral ground states for antiferromagnetic $J_2 > |J_1|/8$. 
Describing a single coplanar  spin spiral by a momentum vector $\vec{q}$ (indicating its direction and pitch), the degenerate
ground-state manifold can be captured by a set of $\vec{q}$ vectors that span a ``spin spiral surface" in momentum space  \cite{Bergman2007}
as illustrated in Fig.~\ref{fig:lattice}.
While these spiral surfaces bear a striking resemblance to Fermi surfaces \cite{Attig2017}, they are considerably more delicate 
objects that can be easily destroyed by small perturbations to the Hamiltonian \eqref{eq:heisenberg-J1J2} (such as further interactions)
or even by fluctuations \cite{Bergman2007,Bernier2008} that will induce an order-by-disorder transition into a simple magnetically ordered state (typically captured by a single $\vec{q}$ vector).
Such a description of the magnetism of A-site spinels in terms of classical local moments has proved sufficient to capture
the physics of the Mn and Co-based spinels \cite{Bergman2007,Lee2008,Savary2011,Gao2017} with local moments $S$=5/2 and $S$=3/2, respectively, while the physics of \FeScS~($S$=2) is dominated by the formation of a spin-orbit coupled local moment \cite{Chen2009,Chen2009b}.

Earlier this year, the synthesis of the first spin-1 A-site spinel has been reported -- \NiRhO, which is found to exhibit no thermal ordering transition 
down to 0.1~K \cite{Chamorro2017}, possibly indicating the formation of a quantum spin liquid ground state.
This motivates us to consider the {\em quantum} version of the minimal exchange model \eqref{eq:heisenberg-J1J2} for spins of arbitrary 
length $S$ in this manuscript and ask whether qualitatively new physics arises in the crossover from the classical to the quantum regime (upon decreasing the spin length). 
We work with a pseudo\-fermion functional renormalization group (pf-FRG) approach \cite{Reuther2010} that has proved capable of handling
competing interactions and emergent spin liquid physics in three-dimensional, frustrated quantum magnets \cite{Iqbal2016,Buessen2016,Iqbal2017} 
and which has recently been generalized to spin-$S$ systems \cite{Baez2016}. 
Our numerical results indicate that a distinct classical to quantum crossover occurs for spin $S=3/2$. 
While the low temperature physics 
is dominated, independent of the spin length $S$, by the formation of spin spiral correlations that manifest themselves in the spin structure factor in the form of clearly discernible spin spiral surfaces (akin to the one shown in the right panel of Fig.~\ref{fig:lattice}), we find that only for systems with spin $S \geq 2$ do these correlations proliferate and give rise to a thermal phase transition into a magnetically ordered ground state. 
For systems with spin $S \leq 1$ we find no indication of a thermal phase transition for the full extent of the spiral regime $J_2/J_1 > 1/8$.
The system with $S = 3/2$ is found to sit precisely at the border with no thermal phase transition occurring in the regime $1/8 < J_2/J_1 \lesssim 0.4$
and a thermal phase transition into a magnetically ordered ground state for $J_2/J_1 \gtrsim 0.4$.
For the spin-1 system of interest in the context of \NiRhO\ these findings support the notion that quantum fluctuations paired with strong geometric frustration can indeed prevent the formation of magnetic ordering and that  
the system remains fluctuating amongst different spin spiral states down 
to the zero temperature. However, when considering a slightly modified exchange model with two distinct types of next-nearest neighbor exchanges
that has been proposed \cite{Chamorro2017} for the tetragonal phase of \NiRhO\ we find that this picture no longer holds. In fact, we find that the modified energetics strongly inhibit the spin spiral fluctuations  
and instead favor the formation of conventional N\'eel order for arbitrary spin length $S$. 
We will return to this point towards the end of the manuscript and discuss how to possibly consolidate these findings 
with the experimental absence of a thermal phase transition.


\noindent
{\em Pseudofermion FRG.--}
To explore the exchange model \eqref{eq:heisenberg-J1J2} we employ the pf-FRG approach \cite{Reuther2010}, 
which recasts the original spin degrees of freedom in terms of auxiliary Abrikosov fermions and then 
applies the well-developed FRG approach of fermionic systems \cite{Wetterich1993, Katanin2004}.
In the language of the original spin model, the pf-FRG approach amounts to a concurrent $1/S$ and $1/N$ expansion 
that allows to faithfully capture conventionally ordered magnetic states (typically favored already in the large-$S$ limit of the expansion) 
and spin liquid states (favored in the alternate large-$N$ limit) 
and is known to become exact in the separate limits of large $S$ \cite{Baez2016} and large $N$ \cite{Buessen2017,Roscher2017}.
With the computational effort scaling quadratically with system size $\mathcal{O}(N_L^2)$ and quartically with the number of frequencies
$\mathcal{O}(N_\omega^4)$, there is a trade-off in choosing larger system sizes versus finer energy/temperature resolution. 
With a focus on the finite-temperature ordering tendencies in the RG flow, we have opted  in our numerical simulations for a 
very finely spaced frequency mesh of 144 frequencies (in a logarithmic spacing) and a system size of $L=10$ lattice bonds 
in every spatial direction (with a total of $N_L=981$ sites) resulting in a total number of 24,219,720 differential equations to be integrated
for every choice of coupling parameters.


\begin{figure}[t]
  \centering
  \includegraphics[width=\linewidth]{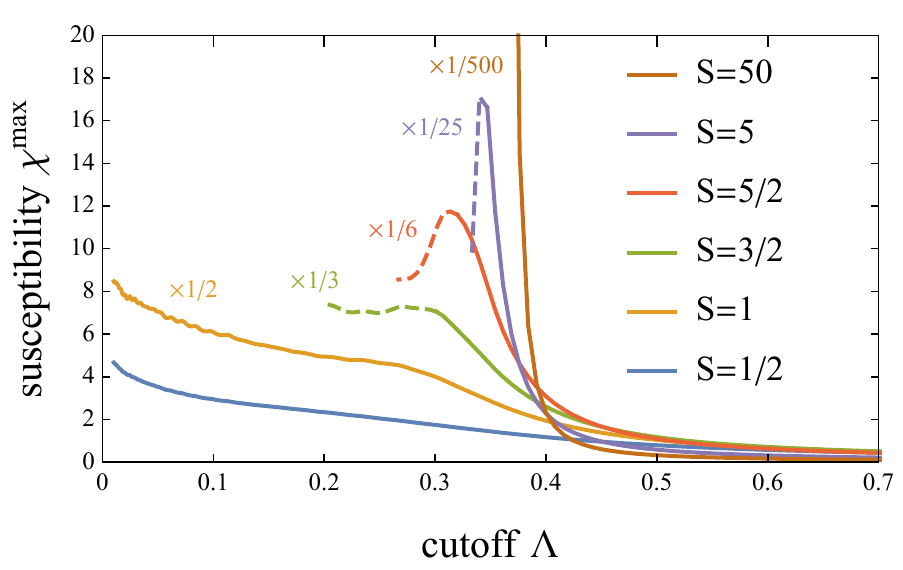}
  \caption{ {\bf Flow of susceptibility} for different spin lengths $S$ for fixed couplings $J_2/|J_1|=0.73$. 
  		The energy scale is normalized by spin length and coupling strength, such that the flow breakdown occurs at similar scales. 
		The susceptibility is always plotted at the momentum-space location where it is maximal. }
  \label{fig:flow_g073}
\end{figure}

\begin{figure}[b]
  \centering
  \includegraphics[width=\linewidth]{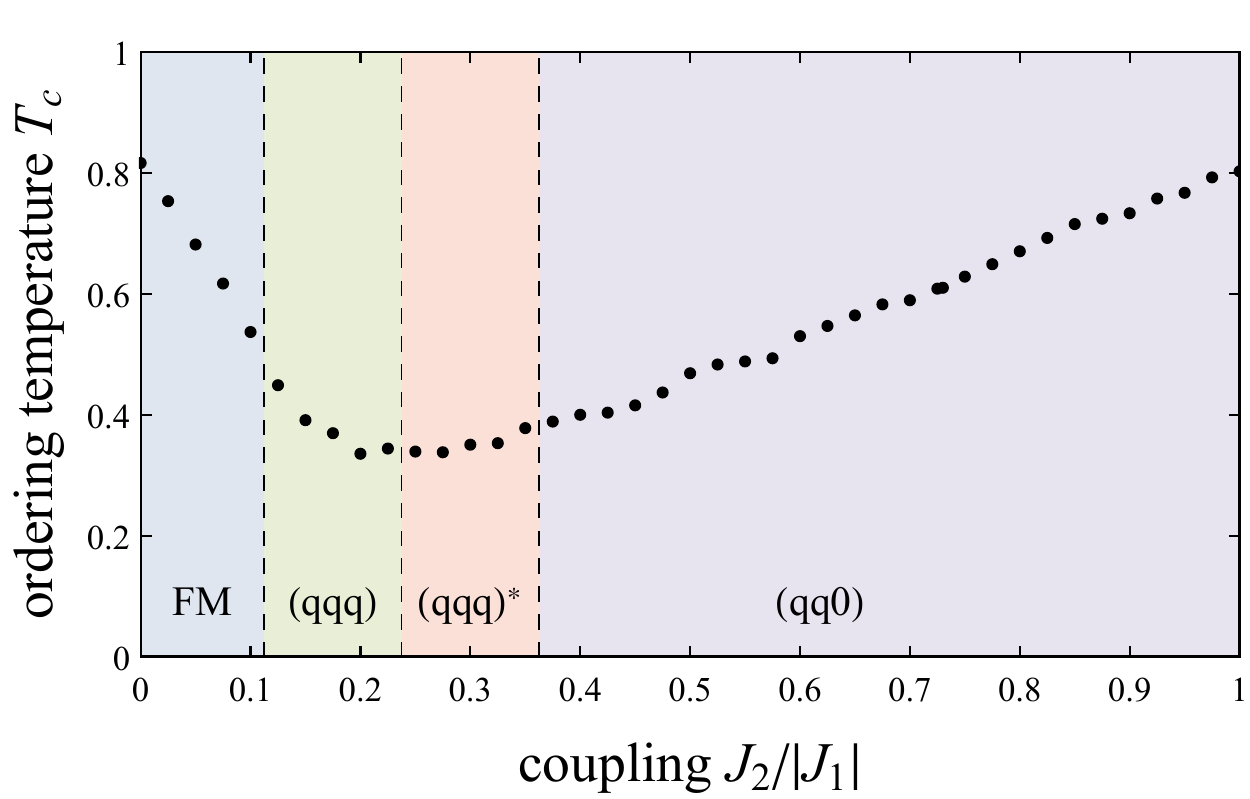}
  \caption{ {\bf Finite-temperature phase diagram}. 
  		Shown is the transition temperature versus the coupling ratio $J_2/|J_1|$ for spin $S=5/2$.
		The background shadings indicates the different types of ground-state order, 
		see the ground-state phase diagram of Fig.~\ref{fig:phasediag}.}
  \label{fig:ordertemp}
\end{figure}

\setcounter{figure}{4}
\begin{figure*}[th]
  \centering
  \includegraphics[width=\linewidth]{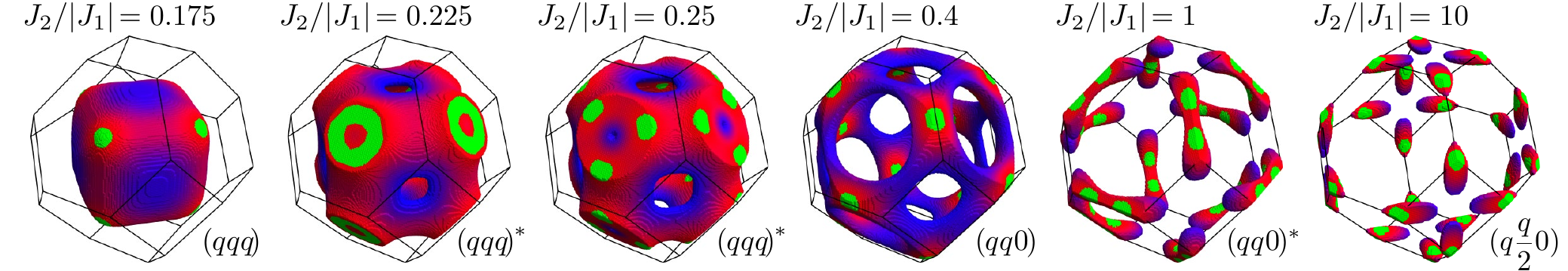}
  \caption{{\bf Spin structure factor of the spin-1 model for varying coupling $\mathbf{J_2/|J_1|}$}. 
			Depicted are the top 20\% of the spin structure factor at frequency cutoff $\Lambda=0$
			with the same color coding applied as in Fig.~\ref{fig:sfbulk_g073_varc}. 
			The spin structure factor shows sharp surface-like features whose evolution with $J_2$ reflects  
			the spin spiral surface found in the ground state of the classical $J_1$-$J_2$ exchange model \cite{Bergman2007}.
			The maxima (indicated in green) describe a sequence of enhanced wavevectors (which characterize the onset of magnetic order for spins $S\geq 3/2$) at
			$(qqq) \to (qqq)^* \to (qq0) \to (qq0)^* \to (q\frac{q}{2}0)$ as $J_2$ is increased.		
			Note that since the maximum of the structure factor is typically hidden inside the finite extent of the depicted manifold
			(see the right panel of Fig.~\ref{fig:sfbulk_g073_varc} for an illustration) we project the maximum radially onto the 
			surface of the manifold.}
  \label{fig:sfBulk_S1}
\end{figure*}

\noindent
{\em Phase diagram.--}
A common starting point for the analysis of a pf-FRG calculation is to plot the magnetic susceptibility as a function of frequency cutoff $\Lambda$
as shown in Fig.~\ref{fig:flow_g073} for the exchange model \eqref{eq:heisenberg-J1J2} at fixed coupling $J_2/|J_1| = 0.73$ (relevant to \NiRhO)
and varying spin length $S$.
For small spins $S$=1/2 and $S$=1 the susceptibility follows a smooth trajectory  down to the lowest temperature and there is no obvious breakdown of the RG flow, which is typically interpreted as the absence of any magnetic ordering transition.
Contrarily, for spins $S$=3/2 and larger the RG flow exhibits a clear breakdown that signals the onset of magnetic order. In fact, what is only a kink in the flow at $S$=3/2 becomes a true divergence in the classical limit ($S$=50). We note that the critical cutoff $\Lambda_c$ \cite{comment-rescale} 
at which the flow breaks down slightly shifts towards larger values for increasing spin length indicating a stronger ordering tendency as one approaches the classical limit.

Identifying the critical cutoff $\Lambda_c$ with a transition temperature $T_c = \Lambda_c \, \pi/2$ \cite{Iqbal2016,Buessen2016} we can map out, for this classical regime, a finite-temperature phase diagram upon varying the ratio $J_2/|J_1|$ at fixed $J_1=-1$ \cite{comment-ferromagnetic}, as illustrated for $S$=5/2 (relevant e.g.~to \MnScS) in Fig.~\ref{fig:ordertemp}. 
Similar to Monte Carlo results \cite{Bergman2007} for the classical exchange model, we find a significant suppression of the transition
temperature for $1/8 \lesssim J_2/|J_1| \lesssim 0.4$, i.e. upon entering the spin spiral regime.

\setcounter{figure}{3}
\begin{figure}[b]
  \centering
  \includegraphics[width=\linewidth]{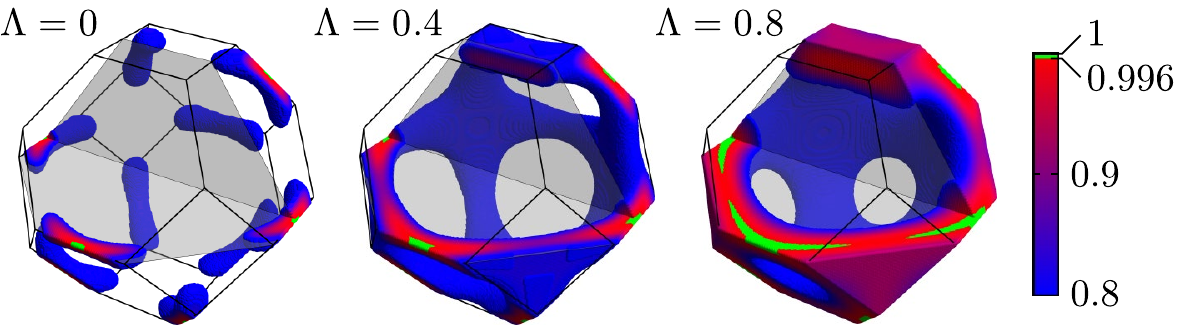}
  \caption{{\bf Evolution of the spin structure factor} with frequency cutoff $\Lambda$ for coupling $J_2/|J_1| = 0.73$ and spin $S=1$. 
  		The colored regions mark the top 20\% of the structure factor. Blue corresponds to 80\% of the maximum value, red to 99.6\%. 
		The top 0.4\% are colored green.}
  \label{fig:sfbulk_g073_varc}
\end{figure}
\setcounter{figure}{5}

To explore the onset of magnetic ordering it is highly instructive to track the evolution of the spin structure factor in the RG flow. 
This is illustrated for the spin-1 model in Fig.~\ref{fig:sfbulk_g073_varc} below where for fixed coupling $J_2/|J_1| = 0.73$ 
we plot the top 20\% of the spin structure factor and the color code reflects the relative strength -- blue is low, red is high, and
green is the top 0.4\%. For large cutoff $\Lambda$ the system fluctuates widely among many different possible 
magnetic orderings.
In the low-temperature, small cutoff regime, however, we find that the features of the spin structure factor sharpen considerably
and become highly reminiscent of the spin spiral surface found   
for the ground state of the classical $J_1$-$J_2$ exchange model \cite{Bergman2007}. 
This is visualized for the spin-1 model for various values of the coupling ratio $J_2/|J_1|$ in Fig.~\ref{fig:sfBulk_S1} above.
Ignoring the coloring scheme for a moment, one sees that the spin structure factor indeed retraces
the spin spiral surface evolving from a spherical object for small $1/8 < J_2/|J_1| \lesssim 0.2$ to an open surface that touches the 
border of the Brillouin zone and forms holes around the $(qqq)$-direction for larger $J_2$ to more line-like objects first around the
$(q q 0)$ direction for $J_2/|J_1| \approx 1$ to two crossing line-like objects in the large $J_2$ limit. 
These observations fall in line with results for the spin structure factor of the classical exchange model obtained from Monte Carlo 
simulations \cite{Bergman2007}.
Here our focus is on further discerning the subset of points within the spiral surface where the structure factor is maximally enhanced,
which provides an indicator of the magnetic ordering that will proliferate in case of a thermal phase transition and 
determine the ground state order.
\begin{figure}[b]
  \centering
  \includegraphics[width=\linewidth]{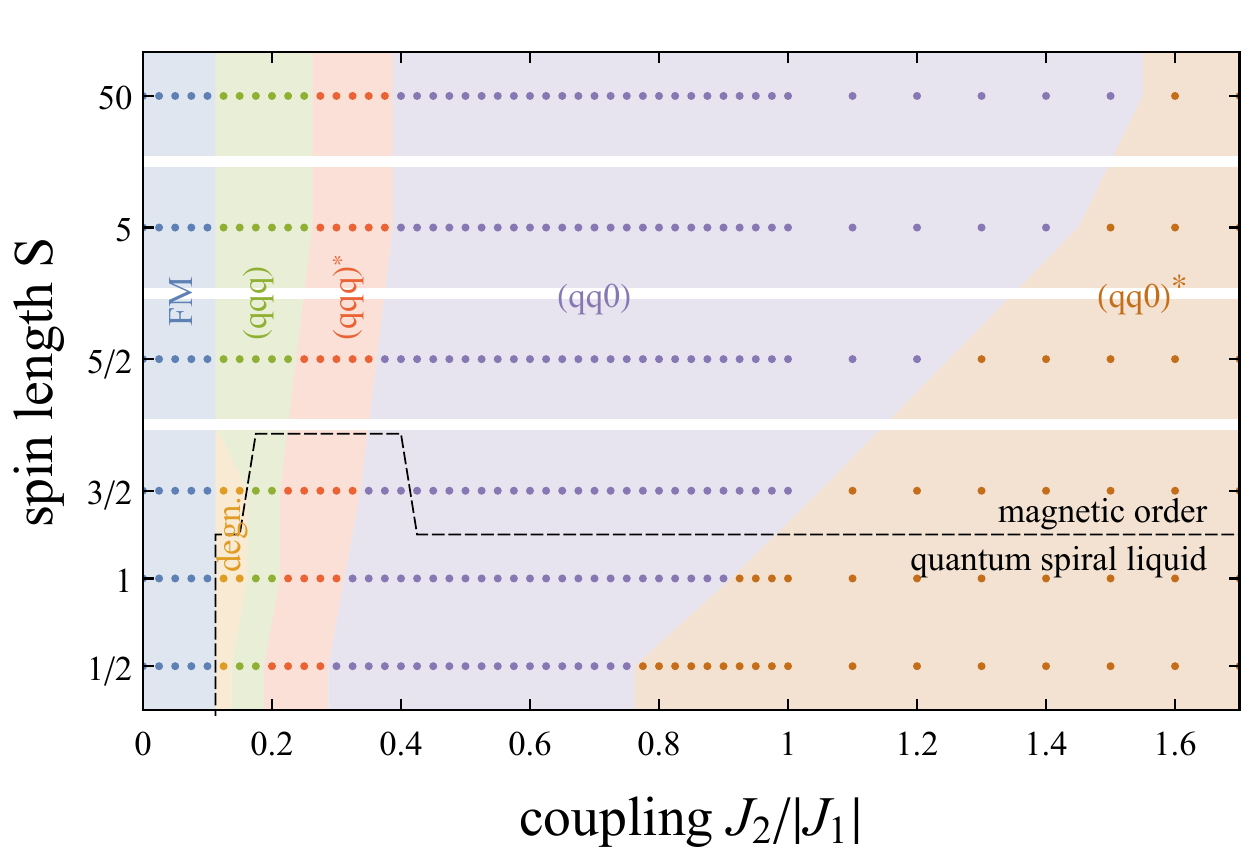}
  \caption{ {\bf Ground state phase diagram}. 
  		The zero-temperature magnetic ordering (indicated by the coloring) as a function of the coupling ratio $J_2/|J_1|$
		and spin lengths varying from the quantum limit $S=1/2$ (bottom) to the classical limit $S=50$ (top).}
  \label{fig:phasediag}
\end{figure}
Tracking these points one finds that beyond the N\'eel / ferromagnetic state for vanishing $J_2$ the preferred ordering momenta go
for increasing $J_2$ through a sequence $(qqq) \to (qqq)^* \to (qq0) \to (qq0)^* \to (q\frac{q}{2}0)$ (where the asterisk 
marks an ordering direction around a high-symmetry direction).

Repeating this analysis for varying spin length $S$ allows us to map out the general ground-state phase diagram of Fig.~\ref{fig:phasediag} 
as a function of both the coupling ratio $J_2/|J_1|$ and spin length $S$. We find that the general evolution of the spiral surface and the sequence 
of incipient ordering momenta do not change upon going from the quantum regime ($S$=1/2) deep into the classical regime ($S$=50) with only
the boundary between the $(qq0)$ and $(qq0)^*$ order showing a noticeable dependence on the spin length $S$.

\noindent
{\em Quantum spiral spin liquids.--}
With the spin structure factor revealing the spiral surface, i.e.~the manifold of approximately degenerate spin spirals at low temperatures, 
we can systematically investigate the effect of quantum fluctuations by varying the spin length $S$.
In the classical limit ($S$=50), the spiral surface determined via the spin structure factor indeed maps
out a manifold of similar size and shape as found in the Luttinger-Tisza calculation \cite{Luttinger1951,Luttinger1946} for the ground state of the classical model (see appendix).
Increasing quantum fluctuations with decreasing spin length $S$, the spiral surfaces become not only more pronounced but systematically expand, similar to the trend observed for increasing the geometric frustration by ramping up $J_2$ in Fig.~\ref{fig:sfBulk_S1}. This expansion can be readily explained by the fact that quantum systems gain more energy from antiferromagnetic fluctuations as opposed to ferromagnetic ones \cite{comment-phaseboundaries}.
The absence of a thermal phase transition (see also Fig.~14 in the appendix) for the low-spin systems with 
$S$=1/2 and $S$=1 points towards the formation of an unconventional ground state. In fact, the system remains fluctuating amongst different spin spiral states
down to zero temperature.
We dub this heavily fluctuating quantum state a {\em quantum spiral spin liquid} and note that this is a decisively different state from the 
topological paramagnet \cite{Wang2015} recently suggested as ground state for the spin-1 model at hand.


\noindent
{\em \NiRhO.--}
Let us finally turn to the spin-1 A-site spinel \NiRhO\ \cite{Chamorro2017}, whose recent synthesis has motivated the
current study of the spin-$S$ $J_1$-$J_2$ exchange model \eqref{eq:heisenberg-J1J2}. 
\NiRhO\ exhibits strong antiferromagnetic couplings with a Curie-Weiss temperature of $\Theta_{\rm CW} \approx -10$~K
and shows no signs of a magnetic ordering transition down to $0.1$~K \cite{Chamorro2017}. 
While one might hope that this makes \NiRhO\ a prime candidate for the spin liquid physics of frustrated spin-1 diamond
antiferromagnets discussed in this manuscript, there are some indications that the exchange model of Eq.~\eqref{eq:heisenberg-J1J2}
needs to be further expanded to truthfully capture the physics of \NiRhO. 
For one, a structural transition of \NiRhO\ around $T\approx 400K$ introduces a tetragonal distortion that requires to 
discriminate between in-plane $J_2^-$ and out of plane $J_2^\perp$ next-nearest neighbor couplings (indicated in Fig.~\ref{fig:lattice}).
Ab initio theory \cite{Chamorro2017} suggests that the relevant coupling strengths for  \NiRhO\
are given by $J_1 = 1$, $J_2^- = 0.73$, $J_2^\perp = -0.91$ with antiferromagnetic $J_1$, $J_2^-$ and ferromagnetic 
$J_2^\perp$. 
If, however, we consider these two distinct types of next-nearest neighbor couplings, we find, 
both in a Luttinger-Tisza calculation for the classical limit as well as in our pf-FRG calculations for all spin $S$, 
a conventional, N\'eel ordered ground state that is accompanied by a finite-temperature transition 
for arbitrarily small tetragonal splitting of the next-nearest neighbor interactions.

\begin{figure}[t]
  \centering
  \includegraphics[width=\linewidth]{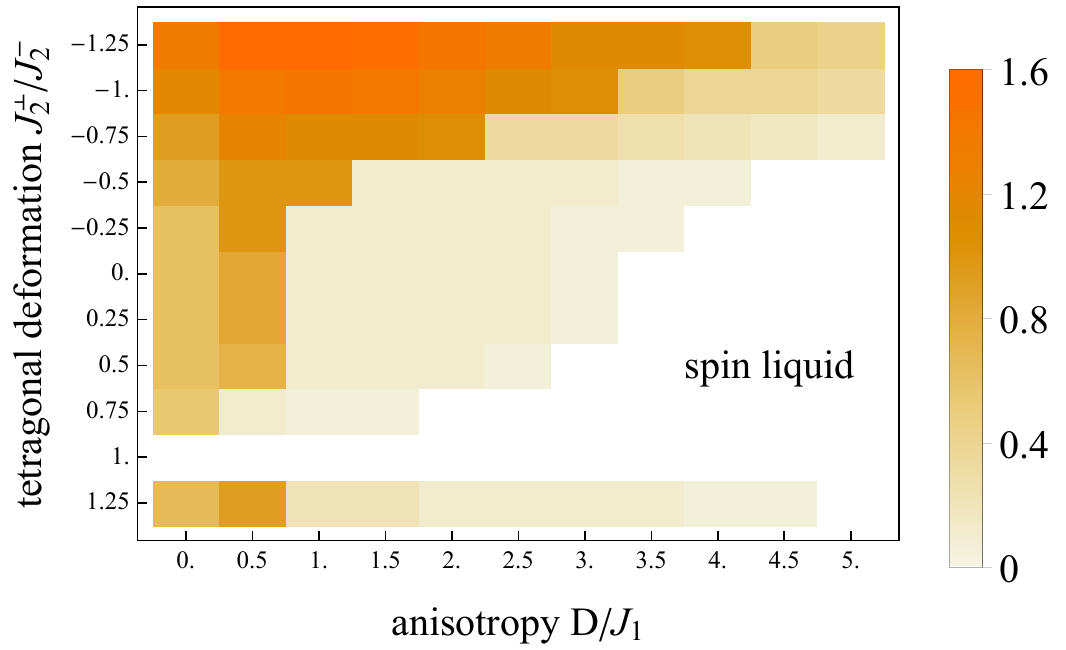}
  \caption{{\bf Effects of a local spin-anisotropy} on the tetragonally deformed diamond lattice.
  		Displayed is the breakdown scale $\Lambda_c$ of the flow. The absence of a breakdown ($\Lambda_c=0$) indicates the absence of magnetic order \cite{comment-rescale2}.}
  \label{fig:tetraDiamondPhaseDiagram}
\end{figure}

One possible way to defy this magnetic ordering tendency in the presence of a tetragonal distortion
is to introduce a local single-ion spin anisotropy term $\sim D\sum_iS^z_iS^z_i$ as a novel source of frustration \cite{Chen2017}.  
Indeed, we find in our pf-FRG calculations \cite{comment-anisotropy} that the latter stabilizes an extended paramagnetic phase
where the system effectively decouples into single sites and thus exhibits a {\em featureless} spin structure 
factor as opposed to the quantum spiral spin liquid discussed above.
For the original $J_1$-$J_2$ model, the spiral spin liquid gives way to a featureless paramagnetic regime
around $D/J_1 \approx 2$ (see appendix), while the magnetic order in the presence of a tetragonal splitting
is more robust and the critical value of the single ion anisotropy quickly rises \cite{comment-anisotropy}. 
The resulting phase diagram is displayed in Fig.~\ref{fig:tetraDiamondPhaseDiagram}, where we plot the RG breakdown scale 
as a function of the strength of the tetragonal deformation and the local spin anisotropy $D/J_1$.
The finite breakdown scale $\Lambda_c$ indicates magnetic order that is stabilized by a tetragonal deformation $J_2^\perp/J_2^- \neq 1$.
For a finite spin anisotropy, however, an extended paramagnetic regime (corresponding to the white regime in Fig.~\ref{fig:tetraDiamondPhaseDiagram} indicating a vanishing $\Lambda_c$) is stabilized.

While for the suggested ab initio  parameters for \NiRhO\ \cite{Chamorro2017} (with $J_2^\perp/J_2^- =-1.25$) we do not see a transition 
into the paramagnet  up to values of $D/J_1 \approx 8$, already a relative small reduction of this coupling ratio (i.e. a smaller  tetragonal splitting) would suffice to explain the absence of magnetic order observed in experiment. 
Looking ahead, it is thus desirable to compare our model calculations with extended ab initio calculations 
that explicitly include the single-ion spin anisotropy. 
Experimentally, neutron diffraction experiments, such as the ones recently undertaken for \MnScS\ \cite{Gao2017}, could
reveal whether the physics of  \NiRhO\ is dominated by spin spiral liquid correlations or trivial paramagnetism.


\begin{acknowledgments}
We thank J. Attig for providing us with the Luttinger-Tisza result for the classical limit of the modified exchange model
for \NiRhO.
This work was partially supported by the DFG within the CRC 1238 (project C02) and Transregio CRC 183 (project A02). 
The numerical simulations were performed on the CHEOPS cluster at RRZK Cologne and the JURECA cluster at the Forschungszentrum Juelich.
J.R. is supported by the Freie Universit\"at Berlin within the Excellence Initiative of the German Research Foundation.
F.L.B. thanks the Bonn-Cologne Graduate School of Physics and Astronomy (BCGS) for support.
\end{acknowledgments}


\bibliography{diamond}

\begin{thebibliography}{37}
\bibitem [{com({\natexlab{a}})}]{comment-rescale}%
  \BibitemOpen
  \bibinfo {note} {Note that we have rescaled all energy scales in the
  manuscript by a factor of $2S\sqrt{J_1^2+J_2^2}$.}\BibitemShut {Stop}%
\bibitem [{com({\natexlab{c}})}]{comment-phaseboundaries}%
  \BibitemOpen
  \bibinfo {note} {This expansion of the spiral surface upon decreasing the
  spin length is also the reason for the shifts of the phase boundaries in the
  ground state phase diagram of Fig. \ref{fig:phasediag}.}\BibitemShut {Stop}%
\bibitem [{com({\natexlab{e}})}]{comment-rescale2}%
  \BibitemOpen
  \bibinfo {note} {Note that energy scales in the plot are renormalized by a
  factor $\sqrt{J_1^2+(J_2^-)^2+(J_2^\perp)^2+D^2}$.}\BibitemShut {Stop}%
\end{thebibliography}

\appendix
\begin{widetext}


\section{Functional renormalization group implementation}
\noindent In this section we supply additional details on the implementation of the functional renormalization group scheme, which is conceptually equivalent to the original formulation of pf-FRG described in \cite{Reuther2010}. 
In principle, the FRG approach provides an exact description of the physical model, equivalent to the functional integral formulation, and does not require any approximations. However, on this formal level, the set of flow equations obtained in FRG is infinitely large. In order to solve this hierarchy of coupled differential equations, one has to make a truncation and keep only a finite number of flow equations. In our calculations we consider only the flow equations for the single-particle and two-particle vertices and neglect those of higher order. The truncation is improved by the Katanin scheme \cite{Katanin2004}, which was shown to be a crucial extension to the formalism in order to capture spin liquid phases \cite{Reuther2010, Buessen2017, Roscher2017}. 
The resulting flow equations for pseudofermions that this work is based on are given in the appendix of Ref. \cite{Reuther2010}.

The structure of the flow equations is such that it contains summations over the entire real-space lattice. To compute these sums numerically, one has to constrain them to a finite system size. While conventional numerical methods typically operate on finite real-space lattices with open or periodic boundary conditions, the pf-FRG scheme naturally uses a different notion of finite size. In pf-FRG calculations the central objects are fermionic interaction vertices. It is therefore straight-forward to include only interactions between fermions that are up to $L$ lattice bonds apart and neglect interactions of fermions further afar. This sets a finite-size scale $L$ but it is does not introduce artificial boundaries. Finite-size effects still exist but they become small already for moderate system sizes. Convergence is reached already at $L\approx 10$ (Fig. \ref{fig:finitesize}).
Particularly for spin spiral configurations the absence of an artificial boundary greatly improves the simulation.
Since the lattice formally does not have a boundary it does not put a limit on the momentum-space resolution of the Brillouin zone and spin spirals can be captured for arbitrary $\vec{q}$-vectors.
\begin{figure*}[h]
  \centering
  \includegraphics[width=0.85\linewidth]{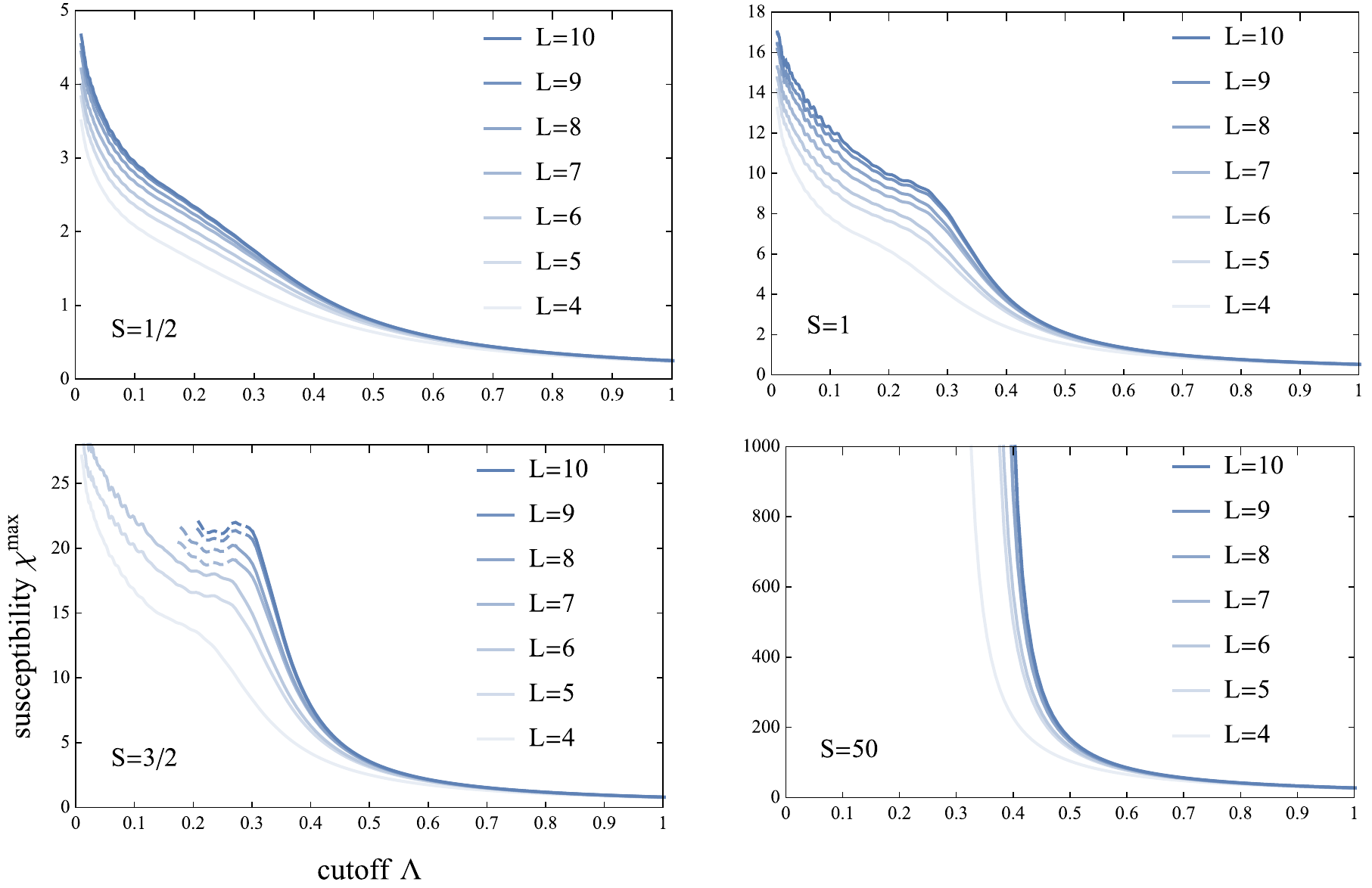}
  \caption{{\bf Finite size effects}. 
  		$J_1$-$J_2$ Heisenberg model at $J_2/|J_1|=0.73$ at different spin lengths $S$ and system sizes $L$. 
		}
  \label{fig:finitesize}
\end{figure*}

A different class of numerical artifacts may also appear in the frequency discretization. At zero temperature where Matsubara frequencies become continuous one has to artificially discretize the frequency space for numerical calculations. The frequency discretization leads to small oscillations in the susceptibility flow (c.f. Fig. \ref{fig:finitesize}) that become weaker as the number of discrete frequencies is increased. To reduce the oscillations we linearly interpolate interaction vertices in frequency space and choose sufficiently many logarithmically spaced frequencies that a physical flow breakdown is not concealed by oscillations.
With these modifications the numerical solution of the flow equations is straight-forward. The flow equations are sufficiently smooth that the Euler scheme produces stable results at reasonable computational costs. 

\clearpage
\newpage


\section{Spin-S consistency checks}
\noindent
{\em Large-S generalization.--}
The generalization of pf-FRG flow equations to larger spins $S\geq1$ is constructed by a substitution of the spin operators by artificial moments $\mathbf{S}_i=\sum_{\kappa=1}^{2S}\mathbf{S}_{i\kappa}$ as discussed in detail in Ref.~\cite{Baez2016}. To ensure projection into the correct subspace of the resulting spin algebra, all spin flavors $\kappa$ must align ferromagnetically. 
This can be energetically enforced by introducing an additional level repulsion term into the Hamiltonian (1) of the main article
\begin{equation}
\label{eq:heisenberg-levelrepulsion}
\mathcal{H}^\prime = \mathcal{H} + A \sum\limits_i\left(\sum\limits_{\kappa=1}^{2S}\mathbf{S}_{i\kappa}\right)^2
\end{equation}
and choosing $A<0$. The same level repulsion term also guarantees single occupation of Abrikosov fermions, even for spin-1/2 systems. 

Numerical data for varying strength of level repulsion $A<0$ are shown in Fig.~\ref{fig:levelrepulsion}.
Note that the depicted susceptibility flows remain largely invariant upon introducing a small, finite level repulsion. 
This indicates that the fermion filling constraints are readily fulfilled in pf-FRG calculations even in the absence of the level repulsion
and the spin-$S$ generalization does indeed hold. 

\begin{figure*}[h]
  \centering
  \includegraphics[width=0.85\linewidth]{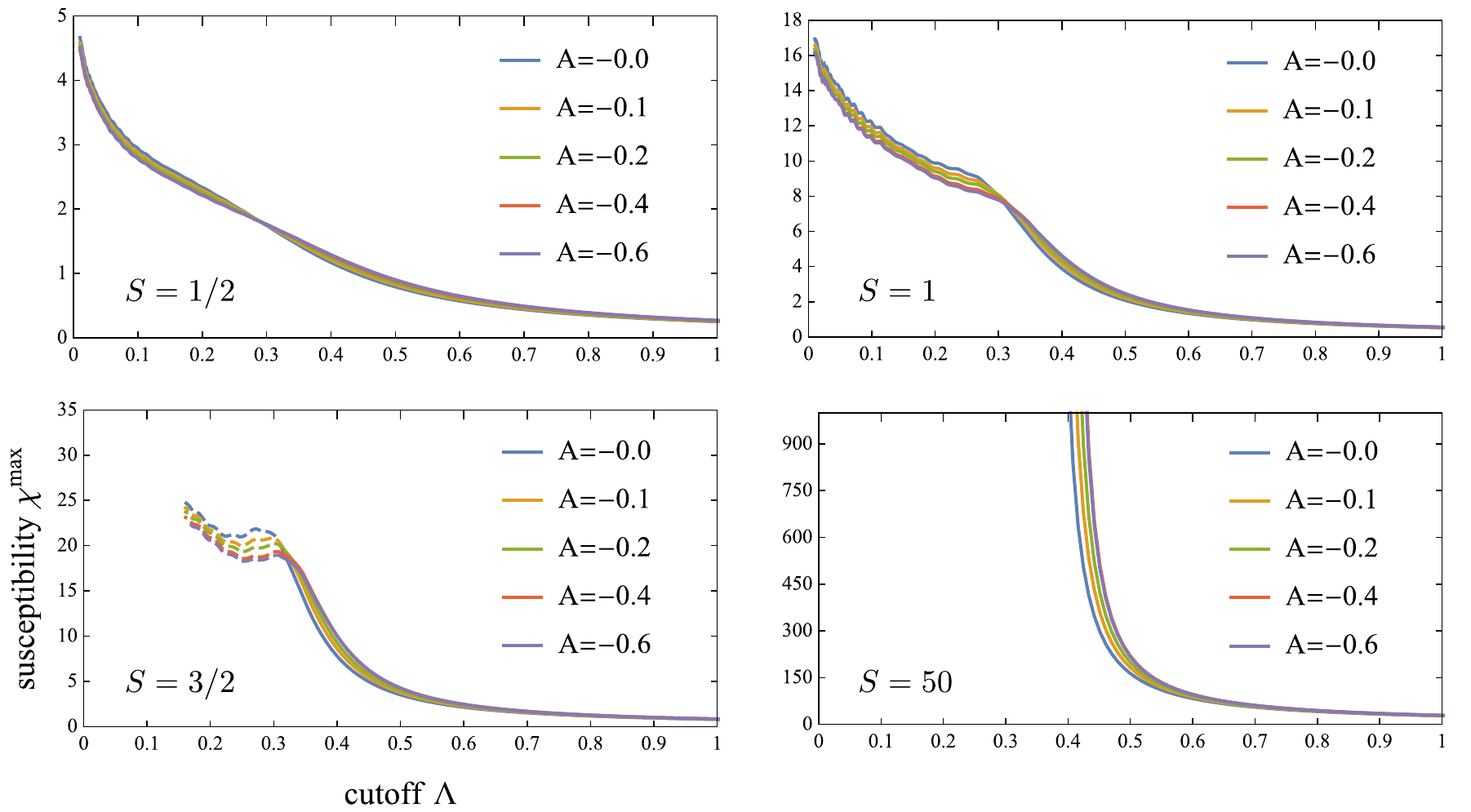}
  \caption{{\bf Level repulsion term in spin-S calculations}. 
  		$J_1$-$J_2$ Heisenberg model at $J_2/|J_1|=0.73$ with finite level repulsion $A<0$. The data has been rescaled to a common energy scale by a factor of $2S\sqrt{J_1^2+J_2^2+A^2}$ such that all data sets coalesce to a single curve. 
		}
  \label{fig:levelrepulsion}
\end{figure*}

\clearpage
\newpage


\section{Local spin anisotropy}
\label{sec:spinanisotropy}

\noindent
To explore possible alternative sources of frustration, we consider a local (single-ion) spin anisotropy in addition to the Heisenberg interactions
\begin{equation}
\label{eq:heisenberg-levelrepulsion}
\mathcal{H}^\prime = \mathcal{H} + D \sum\limits_i S_i^z S_i^z  -  A \sum\limits_i \mathbf{S}_i \mathbf{S}_i \,,
\end{equation}
where $D>0$ parametrizes the strength of the anisotropy. 
Note that within the the pf-FRG scheme for spins $S>1/2$, the anisotropy term may not only drive the system into the $S_i^z=0$ sector, but it could just as well drive the system into the unphysical $\mathbf{S}_i=0$ sector. To constrain the system to the physical sector we need to carefully counterbalance the spin anisotropy term with a level repulsion term (see previous section). 
Since the spin anisotropy term can be recast as $S_i^z S_i^z = \mathbf{S}_i \mathbf{S}_i - S_i^x S_i^x -S_i^y S_i^y$, which includes a contribution of the same form as the level repulsion term, it is apparent that the strength of the level repulsion should be at least $A/D > 1$. 

If we first consider the single-ion limit $D\to\infty$
we find that the in and out-of-plane susceptibilities converge in a range of $3 \lesssim A/D \lesssim 10$ as shown in Fig.~\ref{fig:flow_singleion},
though we note that the out-of-plane susceptibility never vanishes entirely. For large level repulsion strength $A/D \gtrsim 10$ the susceptibilities
start to diverge indicating a breakdown of the pf-FRG framework.

\begin{figure*}[th]
  \centering
  \includegraphics[width=0.45\linewidth]{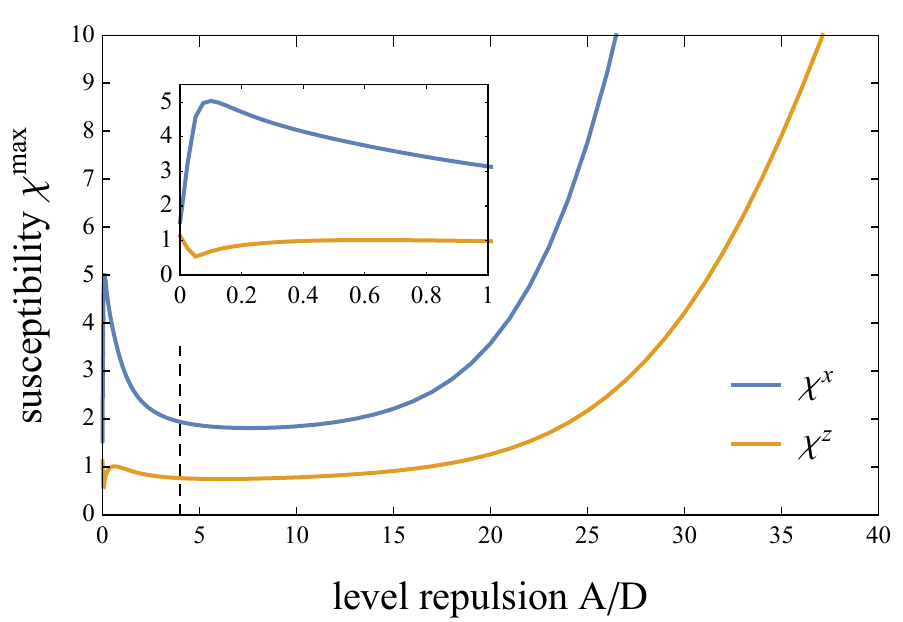}
  \caption{{\bf Local spin anisotropy for a single ion}. 
  		The susceptibility of a single spin-1 moment at zero cutoff plotted for different rations of the anisotropy $D$ and the counter-balancing level repulsion term $A$. At small level repulsion ($A/D\lesssim3$) contributions from the unphysical sector of the Hilbert space dominate as well as for very large level repulsion ($A/D\gtrsim10$). In between there exists a flat plateau where the system is constrained to the physical Hilbert space. }
  \label{fig:flow_singleion}
\end{figure*}

In Figs.~\ref{fig:anisotropy1} and \ref{fig:anisotropy2} we show the in and out-of-plane susceptibility flows for varying strengths of the local spin anisotropy both for the original $J_1$-$J_2$ Heisenberg model (for $J_2/|J_1| = 0.73$) as well as the tetragonal $J_1$-$J_2^-$-$J_2^\perp$ model, respectively. For the spiral spin liquid of the  $J_1$-$J_2$ Heisenberg model we see the expected crossover to the trivial paramagnet of the large $D$ limit at around $D/|J_1| \approx 2$ in accordance with the mean-field estimates of Ref.~\cite{Chen2017}. For the tetragonal model, on the other hand, we do not observe such a transition up to values of $D/|J_1| \approx 8$ as illustrated in Fig.~\ref{fig:anisotropy2}. 

\begin{figure*}[h]
  \centering
  \includegraphics[width=\linewidth]{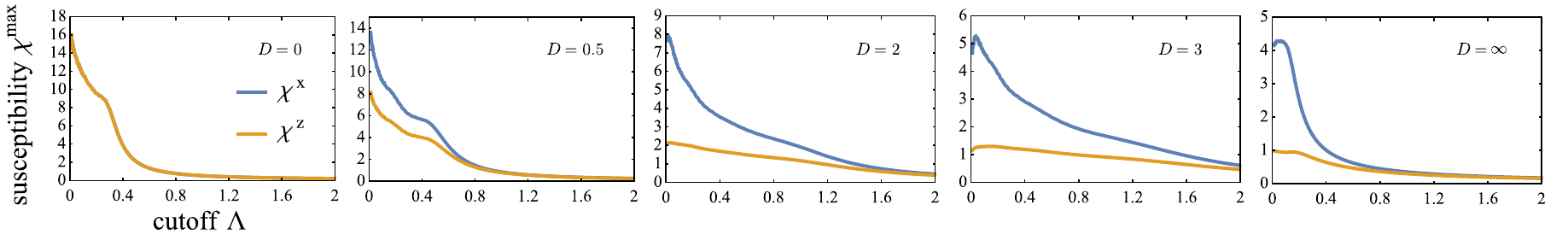}
  \caption{{\bf Spin anisotropy in the $\mathbf{J_1}$-$\mathbf{J_2}$ Heisenberg model} at $J_2/|J_1|=0.73$ and different values for the anisotropy $D$. The level repulsion term $A$ is chosen such that $A/D=4$ for all values of $D$. 
		}
  \label{fig:anisotropy1}
\end{figure*}

\begin{figure*}[h]
  \centering
  \includegraphics[width=\linewidth]{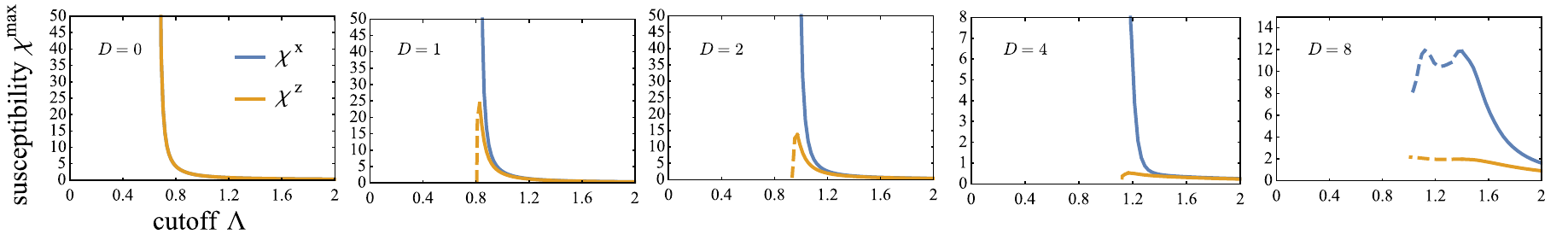}
  \caption{{\bf Spin anisotropy in the $\mathbf{J_1}$-$\mathbf{J_2^-}$-$\mathbf{J_2^\perp}$ model} with coupling constants as suggested in ab initio calculations \cite{Chamorro2017} and an additional spin anisotropy $D$. The level repulsion term $A$ is chosen such that $A/D=4$ for all values of $D$.
		}
  \label{fig:anisotropy2}
\end{figure*}

\clearpage
\newpage


\section{Supplemental data}


\begin{figure*}[th]
  \centering
  \includegraphics[width=\linewidth]{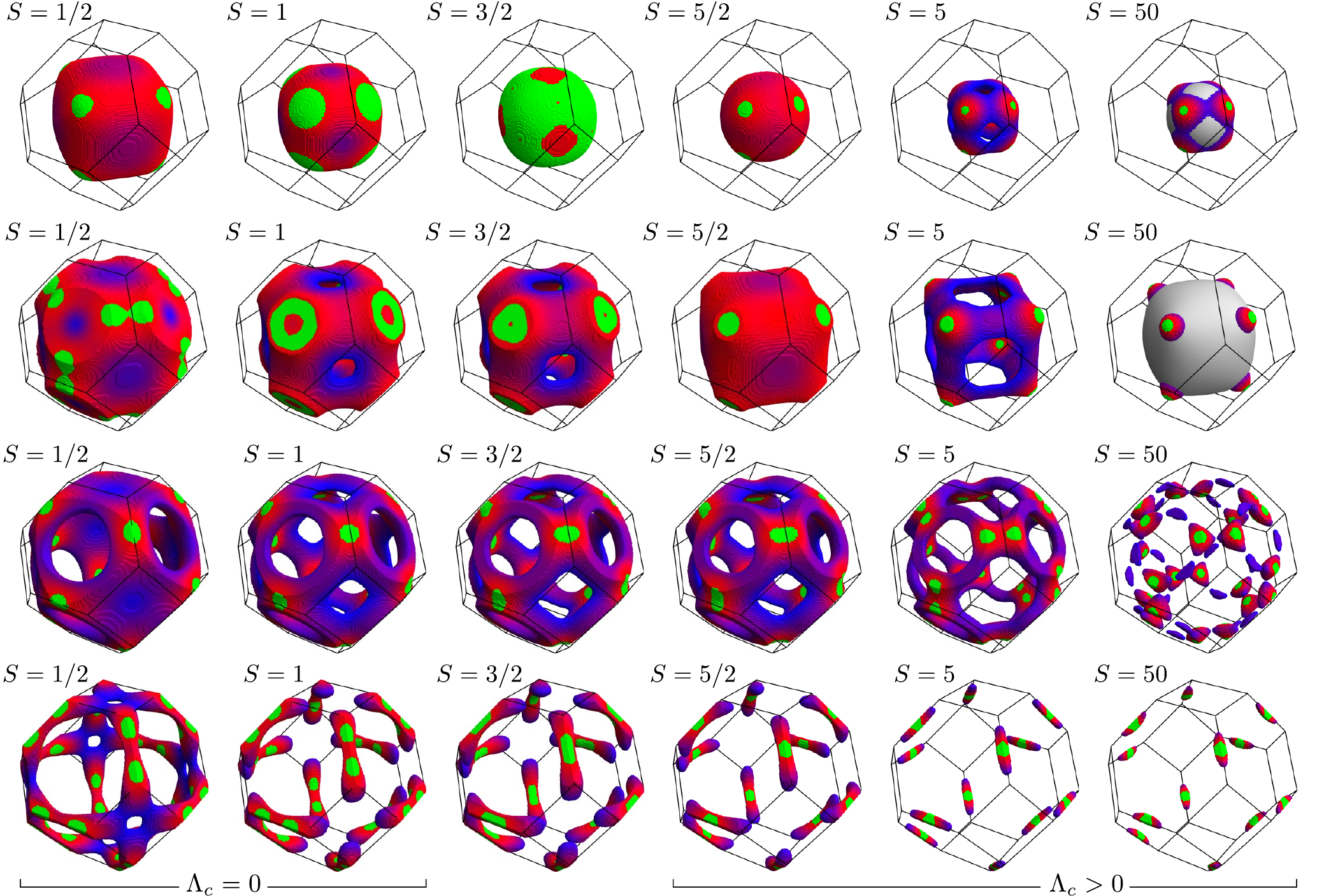}
  \caption{{\bf Effect of quantum fluctuations for varying spin length}. 
		The spin structure factor at the lowest frequency cutoff 
		(either zero or right above the ordering transition) and $J_2/|J_1| = 0.15$ (top row), $J_2/|J_1| = 0.225$ (second row), $J_2/|J_1| = 0.35$ (third row) , or $J_2/|			J_1| = 1$ (bottom row) with the same color coding applied as in Fig.~4 of the main article. 
		The surface-like features reveal nearly degenerate spiral manifolds akin to the spiral surfaces of the ground state of 
		the classical spin model \cite{Bergman2007} (indicated by the grey shaded spheres in the top and the second row).
	Incipient magnetic order for most spin length $S$ is well described by singular points or ring-like shapes in the structure factor. 
	The spin-3/2 system for $J_2/|J_1| = 0.15$ (top row) stands out as the {\em entire} spherical spiral surface remains degenerate down to $\Lambda = 0$. Such a degenerate 		regime upon entering the spin spiral phase, has also been reported in earlier $Sp(N)$ calculations \cite{Bernier2008}.
}  
\end{figure*}


\begin{figure*}[th]
  \centering
  \includegraphics[width=\linewidth]{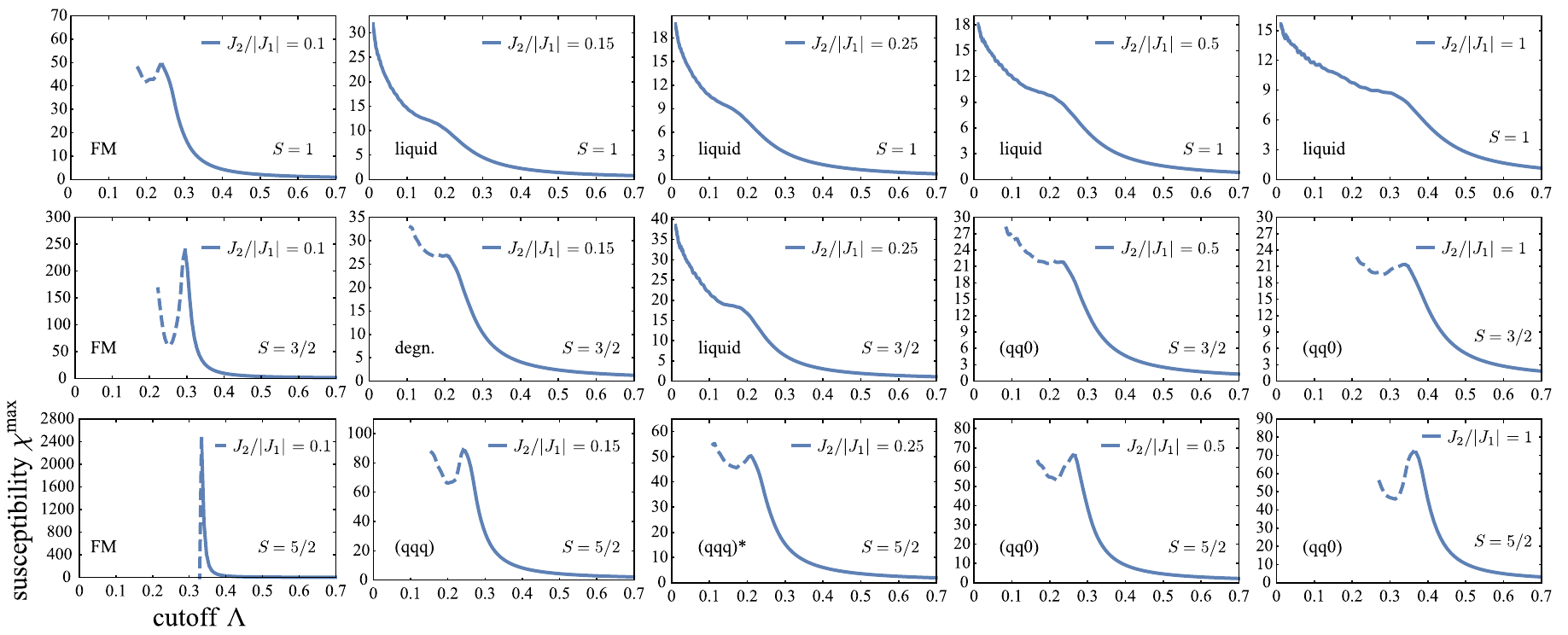}
  \caption{{\bf Flow of the spin susceptibility} for various coupling strengths $J_2/|J_1|$ (columns) and varying $S$ (rows).
  		Shown here is the maximum of the susceptibility versus frequency cutoff $\Lambda$ for the exchange model (1) of the main article
		with $S$=1 (top row), $S$=3/2 (middle row), and $S$=5/2 (bottom row). }
  \label{fig:flow_s1_s15_s2}
\end{figure*}

\end{widetext}
\end{document}